\magnification=\magstep1
\font\bigbfont=cmbx10 scaled\magstep1
\font\bigifont=cmti10 scaled\magstep1
\font\bigrfont=cmr10 scaled\magstep1
\vsize = 23.5 truecm
\hsize = 15.5 truecm
\hoffset = .2truein
\baselineskip = 14 truept
\overfullrule = 0pt
\parskip = 3 truept
\def\frac#1#2{{#1\over#2}}


\topinsert
%
\vskip 1.7 truecm
\endinsert

\centerline{\bigbfont MESOSCOPIC TRANSPORT: THE ELECTRON-GAS SUM RULES}
\centerline{\bigbfont IN A DRIVEN QUANTUM POINT CONTACT}

\vskip 20 truept

\vskip 8 truept
\centerline{
{\bigifont Mukunda P. Das${}^{\dagger}$,
Jagdish S. Thakur${}^{\ddagger}$,}
{\bigrfont and}
{\bigifont Frederick Green${}^{\dagger}$}
}
\vskip 8 truept
\centerline{${\dagger}$\bigrfont Department of Theoretical Physics}
\vskip 2 truept
\centerline{\bigrfont Research School of Physical Sciences and Engineering}
\vskip 2 truept
\centerline{\bigrfont The Australian National University}
\vskip 2 truept
\centerline{\bigrfont Canberra ACT 0200, Australia}
\vskip 8 truept
\centerline{${\ddagger}$\bigrfont Department of Electrical
and Computer Engineering}
\vskip 2 truept
\centerline{\bigrfont Wayne State University}
\vskip 2 truept
\centerline{\bigrfont Detroit, Michigan 48202, USA}
\vskip 2 truept

\vskip 14 truept

\vskip 12 truept
\centerline{\bf 1. INTRODUCTION}
\vskip 12 truept

The electron gas is characterized, above all, by its
response as a correlated many-particle system. In that setting
the conserving sum rules have been thoroughly studied
and exhaustively applied for many years
{[1]}.
The sum rules and their centrality in the electron-gas problem
are widely understood throughout the many-body
community -- if not elsewhere.

At any scale, from the bulk to the near-microscopic,
metallic conduction processes are dominated by the physical
constraints encoded within the electron-gas sum
rules. Thus they apply, with perfect rigor, to mesoscopic
devices such as quantum point contacts. They also
govern {\it nonequilibrium} transport and fluctuations,
as they do the much more familiar linear-response limit.

In this paper we review the structure of the conserving
sum rules in the mesoscopic regime, from the weakest driving
fields to the strongest. 
As compact quantitative expressions of the
microscopic conservation laws,
the sum rules entail a set of straightforward
criteria to be met by proper theoretical descriptions
of metallic-electron behavior.
As we discuss below, these criteria are quite enough
to rule out models of mesoscopic processes that
fail to uphold the conservation laws.

We recall the leading sum rules for mesoscopic conduction.
In particular we apply one of them, the compressibility sum rule, to
strongly driven, open electronic systems. 
Despite their fundamental importance, very little is
known about the role of the sum rules in mesoscopic electronics.
Since the latter is so deeply rooted in the electron gas,
one expects that those relations, as canonical constraints,
carry significant information about the way in which the
processes underlying transport and its associated current noise
are deployed at short scales.

We first discuss the intimate relationship between
the transport response of an open mesoscopic device, connected to
macroscopic metallic leads, and the structure of the microscopic
excitations, or fluctuations, within the device. This relationship
is strongly conditioned by the stabilizing effect of the bounding
leads on the internal dynamics of the device.
Next, we describe how the microscopic form of the
fluctuations in the conductor develops away from equilibrium.
This leads, in particular, to the sum rules for perfect screening
and compressibility
{[1]}.
Then we address the physical implications of the
compressibility sum rule, whose conceptual and practical
import for mesoscopics is on a par with the familiar
fluctuation-dissipation theorem (FDT).

A major, and thus far unexplored, experimental application
of the compressibility sum rule is to mesoscopic systems
with built-in nonuniformity of their conduction bands.
In that context we review some surprising Coulomb effects.
These should be clearly manifested through the behavior of
the electronic compressibility of inhomogenous quantum wires.

\vskip 16 truept
\centerline{\bf 2. BASICS}
\vskip 12 truept

Sum rules express the microscopic conservation laws at the level
of what is measurable in a system. The physical worth of
a theory of conduction can be judged, practically and objectively,
by the degree to which it satisfies them. Approximate descriptions
of an electron system should respect at least the principal ones:

(a) the {\it fluctuation-dissipation relation},
expressing energy conservation,

(b) the {\it perfect-screening sum rule} (gauge invariance), and

(c) the {\it compressibility sum rule} (number conservation).

\noindent
The importance of (a) in mesoscopic transport is well known.
Unfortunately, that of (b) and (c) is not well known at all.
All three rules are equally pivotal; all three must be satisfied.
{\it None} of the three is optional; if even one
of them is violated by a candidate model of electron transport,
that model is manifestly nonviable. Whatever special pleading
may be made for it (for example, some fancied phenomenological simplicity),
it cannot compensate for basic failures of this order.  

Each of the electron-gas sum rules follows a common pattern.
Each is an identity in which a measurable function of the
one-particle distribution for the system is equated to an
integral over a microscopic, {\it two-particle} correlation function.
Let us briefly examine the FDT as a starting point.
In the linear-response limit the FDT provides
an expression for the average power $P$ in steady state,
supplied by the external current generator to the driven system
{[2,3]}:

$$
\eqalignno{
P &= VI = -\lim_{t \to \infty} {\left\{
\int^L_0 E(x) dx {\langle j(x,t)/L \rangle}_I \right\}}
\cr
&\equiv
\lim_{t \to \infty} {\left\{
\int^t_0 \!\!dt' \int^L_0 \!\!dx \int^L_0 \!\!dx'
(-E(x)) {{\langle [{j(x,t)/L}, {j(x',t')/L}] \rangle}_0
         \over k_{\rm B}T} (-E(x'))
\right\}}
\cr
&= {1\over k_{\rm B}T}\int^{\infty}_0 \!\!dt''
\int^L_0 \!\!dx \int^L_0 \!\!dx'
(-E(x)) {\langle [{j(x,t'')/L}, {j(x',0)/L}] \rangle}_0 (-E(x'))
&{(1)}
}
$$

\noindent
where we specialize to one dimension. (This will be
appropriate to a quantum point contact, in which a
quasi-one-dimensional wire is attached to macroscopic metallic leads
{[2]}.) 
The thermal energy is $k_{\rm B}T$, the electromotive force
across the conductor of length $L$
is $V = -\int^L_0 E(x) dx$ and $I$ is the current.
The brackets ${\langle ... \rangle}_I$ represent a trace over the
many-body density matrix for the carriers, subject to transport
(for $I = 0$ it is the equilibrium trace), and
$j(x,t) = -ev(x,t)$ is the particle-flux operator
(to be evaluated in the mean over the channel length,
following the Ramo-Shockley theorem
{[4]}).

The form of Eq. ({1}) is paradigmatic.
On the left we have the experimentally accessible
total power absorbed by the system. On the right,
we have the auto-correlation function
of the {\it microscopic} power-loss density $-E(x)j(x,t)/L$,
which is not directly accessible.
The velocity-velocity correlation
${\langle [v(t), v(0)] \rangle}/k_{\rm B}T$
is a measure of the excursions
of carrier energy taken up by electron-hole pair fluctuations.
As the excited pair states propagate, scatter and decay,
their excess energy dissipates.
Conservation requires that these losses add up to the total
electrical power continuously supplied by the generator.
Equation ({1}) expresses this.

In a uniform system, the local electric field $E = -V/L$
is constant. We obtain the standard Kubo conductance formula
{[5]}

$$
\eqalignno{
G &= {P\over V^2}
= {1\over k_{\rm B}T}\int^{\infty}_0 dt \int^L_0 dx {E(x)\over V}
\int^L_0 dx' {E(x')\over V}
{\langle [j(x,t), j(x',0)] \rangle}/L^2
\cr
&\to {1\over L^2 k_{\rm B}T}\int^{\infty}_0 dt
{\langle [j(t), j(0)] \rangle}.
&{(2)}
}
$$

\noindent
For inhomogeneous systems
the sum rule for conductance has a more general,
but conceptually identical, structure. This is covered
more closely in Ref. {[3]}.

The demonstration of the Kubo formula's
gauge invariance for an {\it open} system
was given in explicit detail by Sols
{[6]}.
Having been established, it furnishes necessary
and sufficient criteria for the system's theoretical
description to be microscopically {\it and} globally conserving.
This contrasts with the status of gauge invariance
within Landauer-B\"uttiker-Imry approaches;
see, for example, the commentary in Ref.
{[7]}
following Eq. (51) therein. 
Eq. ({2}) leads directly to the quantized Landauer
formula as a (very particular) limiting case whose microscopic
proof is free of the customary Landauer phenomenology
{[8]}.
As Eqs. (9)--(11) below demonstrate, the latter badly
compromises both gauge invariance and the compressibility sum rule.

\vskip 16 truept
\centerline{\bf 3. CONSERVATION IN OPEN CONDUCTORS}
\vskip 12 truept

The FDT's far-reaching importance derives from the following.
The conductance $G$, which is directly measurable, can be
computed as a one-particle object within a given transport model.
If, using the same model, one computes the two-body
current auto-correlation (which also determines the measurable
Johnson-Nyquist noise), then it must yield precisely
the same value of $G$. That value may, or may not, turn out
to be a good fit to experiment.
For the model's integrity, however, the essential point is that
the consistency of the FDT has to be guaranteed {\it internally}.
Otherwise the physical basis of such a description
is flawed, since it fails to conserve energy. It would
then be difficult to take such a description seriously.

A similarly fundamental and normative significance attaches
to the compressibility sum rule for open conductors,
which we now derive. Consider a conductor in electrical contact
with a pair of large metallic reservoirs, so that electrons
are freely exchangeable. At equilibrium, the whole assembly is
strictly neutral. We attach a current generator
across the interfaces between the leads and the device,
to drive a current $I$ through the structure;
from the viewpoint of the nonequilibrium carriers
within the conductor, the interfaces appear as an
external source and sink for the current.
 
In this nonequilibrium situation, three boundary
conditions apply without qualification:

\item {$\bullet$} {\it thermodynamic equilibrium of the reservoirs}.
Away from the disturbed device and its interface regions,
the electron
population is equilibrated and stable at any value of $I$.
Thus the local electron density $n(\mu, T)$ in each
lead is unchanged as a function of the equilibrium
chemical potential $\mu$ and bath temperature $T$.

\item {$\bullet$} {\it Charge neutrality of the reservoirs}.
Within a well-defined finite range beyond the interfaces,
strong metallic screening ensures the neutrality of
the reservoir carriers with their ionic background,
once again independently of $I$.

\item {$\bullet$} {\it Charge neutrality of the driven device}.
Since global neutrality and that of the individual reservoir
leads are each preserved, the neutrality of the intervening,
current-bearing conductor is secured for all values of
the current.
(Note: an independent formal  proof of this criterion
comes from the global gauge-invariance theorem for open conductors
{[6]}.)

Suppose that the bounded region in which the electrons are
appreciably disturbed (namely the device with its interfaces)
has volume $\Omega$ and contains $N$ mobile electrons
on average. The electrons are fully compensated
at all times. Hence the neutrality conditions are
equivalent to the statement that

$$
{\delta N\over \delta I} = 0 = {\delta \Omega\over \delta I}
{~~}{\rm for{~}all}{~}I.
\eqno{(3)}
$$

\noindent
In a time-dependent situation, the electron distribution
function is the trace of the local-number quantum
operator $\rho_{\bf k}({\bf r}, t)$ with the
nonequilibrium density matrix in the presence of the current:

$$
f_{\bf k}({\bf r}, t; I) = {\langle \rho_{\bf k}({\bf r}, t) \rangle}_I
$$

\noindent
in which 
${\bf k}$ and ${\bf r}$ are, respectively, the wave-vector and
real-space labels (spin and valley indices are subsumed in ${\bf k}$,
and we assign the effective volume normalization $(2\pi)^{-\nu}$ to it).
It follows that

$$
\int_{\Omega} d{\bf r}
\int {d{\bf k}\over (2\pi)^{\nu}} f_{\bf k}({\bf r}, t; I)
= N = \int_{\Omega} d{\bf r}
\int {d{\bf k}\over (2\pi)^{\nu}} f^{\rm eq}_{\bf k}({\bf r}).
\eqno{(4)}
$$

\noindent
Since absolute neutrality holds, the number of carriers in
the active region cannot change from the equilibrium value
determined by the distribution
$f^{\rm eq} \equiv {\langle \rho \rangle}_{I=0}$.

An immediate result of this identity is that {\it any}
fluctuation, $\delta N$,
in $N$ is determined equally well by the corresponding
fluctuation in $f_{\bf k}({\bf r}, t; I)$ out of equilibrium,
as it is by that in
$f^{\rm eq}_{\bf k}({\bf r})$ at equilibrium. Therefore

$$
\int_{\Omega} d{\bf r} \int {d{\bf k}\over (2\pi)^{\nu}}
\delta f_{\bf k}({\bf r}, t; I)
= \delta N = \int_{\Omega} d{\bf r} \int {d{\bf k}\over (2\pi)^{\nu}}
\delta f^{\rm eq}_{\bf k}({\bf r}).
\eqno{(5)}
$$

\noindent
This establishes the perfect-screening sum rule in its most
general form.

\vskip 16 truept
\centerline{\bf 4. NONEQUILIBRIUM COMPRESSIBILITY SUM RULE}
\vskip 12 truept

We may now vary the population $N$ systematically
with respect to the chemical potential, a
procedure that takes its shape from the response of
the equilibrium state. In the important case of the
free electron gas, we know the resulting
analytical expressions:

$$
\eqalignno{
f^{\rm eq}_{\bf k}({\bf r})
&=
{1\over 1 + \exp[(\varepsilon_{\bf k} + U_0({\bf r}) - \mu)/k_{\rm B}T]};
\cr
{\partial f^{\rm eq}_{\bf k}({\bf r})\over \partial \mu}
&= {1\over k_{\rm B}T}
f^{\rm eq}_{\bf k}({\bf r})(1 - f^{\rm eq}_{\bf k}({\bf r})).
&{(6)}
}
$$

\noindent
Note that we incorporate the static mean-field
potential $U_0({\bf r})$ at equilibrium. This is set up
whenever the system has inhomogeneities built into it.

Recalling that
$\Delta f^{\rm eq} = k_{\rm B}T\partial f^{\rm eq}/\partial \mu$
defines the {\it mean-square thermal fluctuation} of the equilibrium
occupation number
{[1]},
the corresponding nonequilibrium fluctuation
$\Delta f(t)$ must satisfy

$$
\int_{\Omega} d{\bf r} \int {d{\bf k}\over (2\pi)^{\nu}}
\Delta f_{\bf k}({\bf r}, t; I)
= \Delta N = \int_{\Omega} d{\bf r}
\int {d{\bf k}\over (2\pi)^{\nu}} \Delta f^{\rm eq}_{\bf k}({\bf r})
\eqno{(7)}
$$

\noindent
where the mean-square total number fluctuation is
$\Delta N = k_{\rm B}T\partial N/\partial \mu$.
This identity, emerging straight out of the perfect-screening rule,
is the heart of the compressibility sum rule.

What does it mean, physically,
to subject $f_{\bf k}({\bf r}, t; I)$ -- an inherently
nonequilibrium object -- to a variation with respect to
$\mu$, an equilibrium parameter? The answer is straightforward.
Whether at equilibrium or not, the asymptotic density
of carriers in the macroscopic leads remains unchanged.
They sense nothing of the conditions (possibly extreme)
that dominate the driven region. On the other hand,
the density in the leads is directly controlled by the
chemical potential local to each reservoir. This local
value is unique, completely unaffected by the current,
and proper to the {\it local} conduction band.
Changing $\mu$ is equivalent to replacing
the neutral reservoirs at electron density $n(\mu)$, say,
with neutral reservoirs at the new density $n(\mu + \delta \mu)$.
At equilibrium, that is the variation's physical meaning.
Because of perfect screening, the variational procedure
remains well defined even when a current is exciting the sample,
all the while connected to its large stabilizing leads.
Perfect screening confines such a disturbance absolutely
to the finite region $\Omega$. Nothing else is touched.

The compressibility sum rule for an open mesoscopic system
can now be formulated. At equilibrium, the rule states that
the compressibility $\kappa$ for $N$ mobile particles
in volume $\Omega$ is given by
{[1]}

$$
\kappa = {\Omega\over N^2} {\partial N\over \partial \mu}
= {\Omega\over N k_{\rm B}T} {\Delta N\over N}.
\eqno{(8)}
$$
 
\noindent
Note especially that $\kappa$ is {\it and must always remain}
independent of any transport parameter. This distinguishes Eq. (8)
from the fluctuation-dissipation relation and Kubo formula.

The compressibility is the inverse of the stiffness
(bulk modulus) of a many-particle system,
which determines its sound velocity.
Hence, customarily, $\kappa$ has been
investigated through sound-velocity measurements.
The pattern typical of the right-hand side of Eq. ({2})
is repeated here; in analogy with the FDT,
it is determined by an integral (equation ({7})) of the
microscopic electron-hole pair correlation,
whose static limit is $-\partial f^{\rm eq}/\partial \mu$.
For a careful exposition, see Ref.
{[1]}.

It is clear from Eqs. ({3}), ({4}) and ({7})
that, since all the quantities on the right-hand side
of the compressibility sum rule are independent of $I$,

\item {$\bullet$} {\it the compressibility of an open mesoscopic
conductor is strictly invariant under transport}.

\noindent
This is the principal result that we wish to recall.
It articulates the conservation of particle number,
just as perfect screening expresses the
general gauge invariance of a conducting system.

\vskip 16 truept
\centerline{\bf 5. APPLICATIONS}
\vskip 12 truept

\centerline{\it a. Violations of the compressibility sum rule}
\vskip 8 truept 

We are ready to revisit and judge some theoretical
assumptions that have gained currency in recent
mesoscopic research
{[9]}.
Though plausible at first blush, they turn out to
vitiate the sum-rule structure of any transport model
relying on them.
In place of the gauge-invariant and microscopically
canonical prescription of current-driven transport
presented above, let us instead posit {\it ad hoc}
that the current in a narrow mesoscopic conductor
is sustained purely by a difference of chemical potentials
between an upstream and a downstream electron reservoir
{[7],[9],[10]}.

Assume, for argument, that the density differential
$n(\mu_{\rm up}) - n(\mu_{\rm dn})$ induced
across the sample by the upstream and downstream
chemical potentials, $\mu_{\rm up}$ and $\mu_{\rm dn}$,
causes a diffusive-like current. Assume too that the electromotive
force measured across the interfaces is
$V = -(\mu_{\rm dn} - \mu_{\rm up})/e$
(and note that it makes no difference at all to the generality of the
argument, whether $V$ is static or time-dependent;
cf Eqs. (4) \& (5)).

The density profile along the conductor
will be some function $n(\mu_{\rm up} - eV(x))$, taking the
boundary values $n(\mu_{\rm up})$ and $n(\mu_{\rm dn})$
at the ends of the sample.
According to such an account, in the linear-response
limit the total number of carriers in the active structure
of length $L$ changes, under transport, by an amount

\vskip 8 truept
$$
\eqalignno{
N(V) - N(0)
&= \int^L_0 [n(\mu_{\rm up} - eV(x)) - n(\mu_{\rm up})] dx
\cr
&\to \int^L_0 {dn\over d\mu_{\rm up}} (-eV(x)) dx
\cr
&\sim - \Delta N(0) {eV\over 2k_{\rm B}T}.
&{(9)}
}
$$

\noindent
The result is patently counter to Eq. ({4}), and consequently
also breaks the compressibility sum rule. 

If the mooted pseudo-diffusive arrangement is not to violate
the mandatory invariance of total particle number for any driven,
necessarily neutral, mesoscopic structure, the nominal density difference
$n(\mu_{\rm up}) - n(\mu_{\rm dn})$ which causes the current
(and so the current itself) cannot be accorded a
standard physical meaning. This makes any conclusions
drawn from such a picture hard -- if not impossible -- to interpret
in a rational way.

The only escape from that unpalatable outcome is
to allow effectively arbitrary inflows and outflows
of charge between the device and its leads.
In the event, the device cannot stay neutral under transport;
there is {\it always} an uncompensated excess (or deficit) of carriers.
Such a dilution of the perfect screening sum rule
immediately destroys gauge invariance
{[6]}.
In any case, one or more of the sum rules is countermanded.
Transport arguments based on this scenario are unphysical.

At this stage we have addressed only a single-electron view
of the compressibility within pseudo-diffusive transport.
However, at the level of the electron-hole pair correlations,
the sum-rule violation is far worse (if that is possible).
To make the point we take a representative pseudo-diffusive
fluctuation analysis, that of Martin and Landauer
{[10]}.
In their theory, the current-current correlation function
is synthesized from the set of all possible quantum-transmission
outcomes that involve a single electron scattering off the core
region of a one-dimensional quantum point contact.
Its (constant) transmission probability is taken to be ${\cal T}$,
where $0 \leq {\cal T} \leq 1$.
Applying Eqs. (2.6)--(2.15) of Ref. {[10]},
the total one-electron ``fluctuation'', say ${\rm d} N$, can be obtained.
This determines every quantity of experimental interest.
We apply it now to the compressibility within the model.

Before doing so, it is vital to distinguish ${\rm d} N$ from the
microscopically prescribed number fluctuation

$$
\delta N \sim (-\partial f^{\rm eq}/\partial \mu)\delta \mu
$$

\noindent
appearing in the perfect-screening sum rule, Eq. (5).
$\delta N$ represents the {\it intrinsically correlated
electron-hole excitations}: the conservation laws that govern
electron-hole pair processes means that these are
irreducible two-body objects
{[1]}.
On the other hand, ${\rm d} N$ is strictly constructed as
a sum of one-body quantities.

The model at hand, and its closely related cousins
{[7],[9]}
take the single-carrier ${\rm d} N$ -- not the two-body
$\delta N$ -- as the fundamental fluctuation. Therefore
the true pair correlations, crucial to the fluctuation structure
of the electron gas, are missing. 
Here the tacit assumption is, instead,
that a fermionic particle-hole pair is always reducible to
two kinematically uncorrelated, quite independent,
single-particle excitation factors.
Such an assumption is entirely contrary to basic Fermi-liquid theory
{[1],[3]}.

Following construction of the single-electron
${\rm d} N$ after Ref. {[10]},
we arrive at the mean-square expectation $\Delta N$
over the quantum point contact:

$$
\eqalignno{
\Delta N
&\equiv
{\langle ({\rm d} N)^2 \!-\! {\langle {\rm d} N \rangle}^2 \rangle}
= L{n(\mu_{\rm up})\over 2\varepsilon_{\rm F}}
{\left[ {\cal T}^2 k_{\rm B}T +
{\cal T}(1 \!-\! {\cal T}){{\mu_{\rm up} \!-\! \mu_{\rm dn}}\over 2}
{\rm coth}
{\left( {{\mu_{\rm up} \!-\! \mu_{\rm dn}}\over 2k_{\rm B}T} \right)}
\right]}
\cr
&= N{k_{\rm B}T\over 2\varepsilon_{\rm F}}
{\left[ {\cal T} + {{\cal T}(1 - {\cal T})\over 3}
{\left( {eV\over 2k_{\rm B}T} \right)}^2
+ {\cal O}{\Bigl( (eV/2k_{\rm B}T)^4 \Bigr)}
\right]},
&{(10)}
}
$$

\noindent
where $\varepsilon_{\rm F}$ is the Fermi energy
at density $n(\mu_{\rm up})$ in the uniform wire
(the Martin-Landauer model works only in
the degenerate-electron limit
$\varepsilon_{\rm F} \gg k_{\rm B}T$).
To leading order in the voltage, the
theory duly predicts the {\it transport-dependent} ratio

$$
{\Delta N\over N}
= {\left[ {\Delta N\over N} \right]}_{\rm CSR}
{\left[ {\cal T} + {{\cal T}(1 - {\cal T})\over 3}
{\left( {eV\over 2k_{\rm B}T} \right)}^2 \right]}
\eqno{(11)}
$$

\noindent
in which

$$
{\left[ {\Delta N\over N} \right]}_{\rm CSR}
= {k_{\rm B}T\over 2\varepsilon_{\rm F}}
$$

\noindent
is the equilibrium ratio of the
total mean-square number fluctuation to total carrier number
as it appears in the transport-invariant
compressibility sum rule, Eq. (8).

Equation (11) clearly contradicts the invariance of Eq. (8). 
{\it Even at equilibrium}, the total mean-square fluctuation
$\Delta N$ in pseudo-diffusive theories depends on the transport
parameter ${\cal T}$. In the strong backscattering limit
it will vanish with ${\cal T}$.

The electronic compressibility cannot depend on any transport parameter.
As we saw in the previous section, $\kappa$ is an equilibrium
property completely insensitive to external sources
of elastic scattering (such as potential barriers)
which determine the transmission fraction ${\cal T}$.
Thus Eq. (11) fails to recover the physically required
compressibility at equilibrium, much less away from it. Nor can one
appeal to Coulomb effects here, since this spurious result
is derived for noninteracting one-dimensional conduction electrons.

\vskip 8 truept 
\centerline{\it b. Electronic compressibility
in nonuniform quantum channels}
\vskip 8 truept 

We turn our attention to some unexplored mesoscopic implications
of the compressibility sum rule.
When the conduction band of a quantum point contact, or similar
device, differs from that of its bulk leads, there will be
contact potentials at the interfaces.
In Eq. ({6}) the mean-field potential $U_0({\bf r})$,
which vanishes when there is no band mismatch between
device and leads, becomes a nontrivial function of the
electron density. The complete variation of
$f^{\rm eq}_{\bf k}({\bf r})$ is

$$
{\delta f^{\rm eq}_{\bf k}({\bf r})\over \delta \mu}
= {\left( 1 - {\delta U_0({\bf r})\over \delta \mu} \right)}
{\left. {\partial f^{\rm eq}_{\bf k}({\bf r})\over
         \partial \mu} \right|}_{U_0}.  
\eqno{(12)}
$$

\noindent
Since

$$
{\delta n({\bf r})\over \delta \mu}
= \int {d{\bf k}\over (2\pi)^{\nu}}
{\delta f^{\rm eq}_{\bf k}({\bf r})\over \delta \mu},
$$

\noindent
the implicit Eq. (12) can be solved by
integrating over ${\bf k}$ on both sides, and
feeding back into it the leading right-hand
factor in closed form. The result is

$$
{\delta f^{\rm eq}_{\bf k}({\bf r})\over \delta \mu}
= { 1\over {\displaystyle 1 + {dU_0\over dn}
{\partial n({\bf r})\over \partial \mu}} }
{\partial f^{\rm eq}_{\bf k}({\bf r})\over \partial \mu}.
\eqno{(13)}
$$

An easy case to analyze is a uniform wire. Then $U_0$ is
constant over most of the wire's length, except in the
neighborhood of the interfaces where the band mismatch
forces $U_0$ to vary. The leading factor on the right-hand
side of Eq. (13) is nearly constant everywhere.
Denote it by $\gamma$. It appears as an additional
factor in the total compressibility for our inhomogeneous system:

$$
\kappa \equiv {\Omega\over N^2} {\delta N\over \delta \mu}
= {\Omega\over N k_{\rm B}T} {\gamma \Delta N\over N}.
\eqno{(14)}
$$

\noindent
The self-screening response of the contact potential
substantially modifies
the compressibility that would otherwise be observed in the absence
of band inhomogeneity. Such effects are particularly strong
in III-V heterojunction quantum-well structures
{[3]}. 

We state the corollary of the perfect-screening Eq. ({5})
in this situation. It is the nonequilibrium compressibility sum
rule for inhomogeneous mesoscopic systems (see Ref.
{[3]}
for its mathematical grounding).

\item {$\bullet$} {\it The compressibility of a nonuniform
mesoscopic conductor, driven at any current, is invariant
and its form is given by Eq. ({14}).}

\noindent
This is a surprising result, for it asserts that, even with the
large changes in the carrier distribution of a device subjected
to high current, there is {\it no} change at all in its
electronic compressibility. That quantity
is fixed, once and for all, by the electrostatic response of the
contact potential at equilibrium.

The Coulomb-induced suppression of compressibility
correlates closely with the analogous suppression
theoretically anticipated for high-field current noise
in heterojunction based, quasi-two-dimensional wires
{[3]}.
The nonequilibrium excess noise of such a structure
is accessible
{[11]}.
It could be systematically measured,
and direct comparison made with studies of the
static compressibility in the same sample.

\vskip 16 truept
\centerline{\bf 5. SUMMARY}
\vskip 12 truept

The main result in this work is the ``rigidity''
of the electronic compressibility in a mesoscopic wire,
in the sense that it cannot respond {\it in any way} to the
internal Coulomb effects that alter the nonequilibrium
arrangements inside the conductor.
That rigidity, coming from the utter dominance of the
boundary conditions in mesoscopic transport,
is striking evidence of the power and reach of the conservation laws.
Novel experiments in this context, especially the comparative
examination of static and fluctuation responses,
would therefore be of interest.

We recapitulate our findings. The conserving sum rules are
universal constraints that apply to any system of mobile electrons.
They express the basic conservation laws through a set of relations
between, on the one hand, one-body properties that are
measurable (conductance, particle number, compressibility)
and, on  the other hand, expectation values of 
microscopically calculable, two-body correlation functions.

Sum rules embody the unified origin of single- and many-particle
behavior in the electron gas.  
We have shown that the principal sum rules for
{\it energy dissipation},
{\it charge neutrality} and {\it number conservation}
are core properties of a mesoscopic conductor. Their
consequences are directly observable in the laboratory.
A key quantity for experimental investigation is the compressibility.

The implications of the compressibility sum rule have not
previously been studied in the mesoscopic setting.
In sharp contrast with its crucial and long-established
role in electron-gas physics
{[1]},
this property has not had any attention paid to
it by more popularly accepted phenomenologies
of mesoscopic transport and noise
{[7],[9],[10]}.
It is not so astonishing, then, to find that the
compressibility sum rule is violated by such approaches. 

Through the gauge invariance of the boundary conditions
and their overriding influence on internal carrier dynamics,
it may indeed be said that the conservation laws and their
leading sum rules instill the very meaning of ``mesoscopic'' conduction.
Therefore a microscopically credible description of mesoscopics
must satisfy {\it all} these rules, not just a subset,
else the description is faulty.
Its reliability is void of any guarantee -- notwithstanding
any perceived empirical or cosmetic merits.

We appear to be enjoying a phase of development in mesoscopics
wherein heuristically dominated theory-making is much
to be preferred over labor-intensive, assiduously orthodox
and microscopically correct analyses of electron transport.
In such a climate, Sir Francis Bacon's old warning against subjective
confections of knowledge remains apt today:

\vskip 8 truept
\noindent
{\it God forbid that we should give out a dream of our
imagination for a pattern of the world.}
\vskip 8 truept

\vskip 28 truept
\centerline{\bf REFERENCES}
\vskip 12 truept

\item{[1]}
D. Pines and P. Nozi\`eres, {\it The Theory of Quantum Liquids}
(Benjamin, New York, 1966).

\item{[2]}
W. Magnus and W. Schoenmaker, {\it Quantum Transport in
Sub-micron Devices: A Theoretical Introduction} (Springer, Berlin, 2002).

\item{[3]}
F. Green and M. P. Das, {\it J. Phys.: Condens. Matter} {\bf 12} 5251, 2000.

\item{[4]}
C. J. Stanton, PhD Thesis, Cornell University, unpublished (1986).

\item{[5]}
J. M. Ziman, {\it Models of Disorder}
(Cambridge University Press, Cambridge, 1979), Ch 10.

\item{[6]}
F. Sols, {\it Phys. Rev. Lett.} {\bf 67}, 2874 (1991).

\item{[7]}
Ya. M. Blanter and M. B\"uttiker, {\it Phys. Rep.} {\bf 336}, 1 (2000).

\item{[8]}
M. P. Das and F. Green, {\it J. Phys.: Condens. Matter} {\bf 15}, L687 (2003).

\item{[9]}
Y. Imry and R. Landauer, {\it Rev. Mod. Phys.} {\bf 71}, S306 (1999).

\item{[10]}
Th. Martin and R. Landauer, {\it Phys. Rev. B} {\bf 45}, 1742 (1992).

\item{[11]}
M. Reznikov {\it et al}., {\it Phys. Rev. Lett.} {\bf 75}, 3340 (1995).

\end{document}